\documentclass[10pt,twocolumn,superscriptaddress,showpacs,preprintnumbers,amsmath,amssymb,prb,aps]{revtex4}
\usepackage{graphicx}

\begin{document}


\title{Two-gap superconductivity in Lu$_2$Fe$_3$Si$_5$: a transverse-field muon spin rotation study}

\author{P.K. Biswas}
\email[]{P.K.Biswas@warwick.ac.uk}
\affiliation{Physics Department, University of Warwick, Coventry, CV4 7AL, United Kingdom}

\author{G. Balakrishnan}
\affiliation{Physics Department, University of Warwick, Coventry, CV4 7AL, United Kingdom}

\author{D.McK. Paul}
\affiliation{Physics Department, University of Warwick, Coventry, CV4 7AL, United Kingdom}

\author{M.R. Lees}
\affiliation{Physics Department, University of Warwick, Coventry, CV4 7AL, United Kingdom}

\author{A.D. Hillier}
\affiliation{ISIS Facility, Rutherford Appleton Laboratory, Chilton, Oxfordshire, OX11 0QX, U.K.}

\date{\today}

\begin{abstract}
The superconducting properties of Lu$_2$Fe$_3$Si$_5$ with $T_c=6.1$~K have been investigated using low-temperature transverse-field muon spin rotation ($\mu$SR) and specific heat measurements.  The magnetic penetration depth at zero temperature, $\lambda\left(0\right)$, is $353(1)$~nm. However, the temperature dependence of the magnetic penetration depth, $\lambda\left(T\right)$ is consistent with a two gap $s+s$-wave model. Low-temperature specific heat measurements on the same sample also show evidence of two distinct superconducting gaps.
\end{abstract}

\pacs{76.75.+i, 74.70.Ad, 74.25.Ha}

\maketitle

The discovery of superconductivity in MgB$_2$ with a $T_c\sim39$ K~\cite{Nagamatsu} has generated a great deal of interest in superconducting materials containing light elements such as B, C, and Si. Among these materials, the ternary-iron silicide superconductors R$_2$Fe$_3$Si$_5$ with R = Lu, Y, or Sc are particular noteworthy due to the presence of iron.~\cite{Braun,Segre} Lu$_2$Fe$_3$Si$_5$ is the most interesting of the ternary-iron silicide superconductors because of its high superconducting transition temperature ($T_c = 6.1$~K), large upper critical field ($\mu_{0}H_{c2} =  6$~T)~\cite{Braun,Umarji} and unconventional superconducting properties. Recently, a detailed study of the low-temperature specific heat on a single crystal of Lu$_2$Fe$_3$Si$_5$ revealed a two-gap superconductivity similar to that seen in MgB$_2$.~\cite{Nakajima}

Lu$_2$Fe$_3$Si$_5$ has a tetragonal Sc$_2$Fe$_3$Si$_5$-type structure (space group $P4/mnc$) consisting of quasi one-dimensional iron chains along the $c$ axis and quasi two-dimensional iron squares parallel to the basal plane.~\cite{Chabot}

Muon spin rotation ($\mu$SR) is a ideal probe with which to study the mixed state of type-II superconductors as it provides microscopic information of the local field distribution within the bulk of the sample. It has often been used to measure the temperature dependence of the London magnetic penetration depth, $\lambda$, in the vortex state.~\cite{Sonier, Brandt} The temperature and field dependence of $\lambda$ can provide information on the nature of the superconducting gap.

Here we have investigated the unusual superconducting properties of Lu$_2$Fe$_3$Si$_5$ by carrying out low-temperature $\mu$SR measurements on a polycrystalline sample. We show that the temperature dependence of $\lambda$ can be well described using a two-gap $s+s$-wave model. The magnetic penetration depth at $T=0$~K is estimated to be $\lambda\left(0\right)=353(1)$~nm. We also study the low-temperature specific heat of Lu$_2$Fe$_3$Si$_5$ in order to support the validity of the two-gap model. We compare these results with published data for the R$_2$Fe$_3$Si$_5$ system.

\begin{figure}[tb]
\begin{center}
\includegraphics[width=0.9\columnwidth]{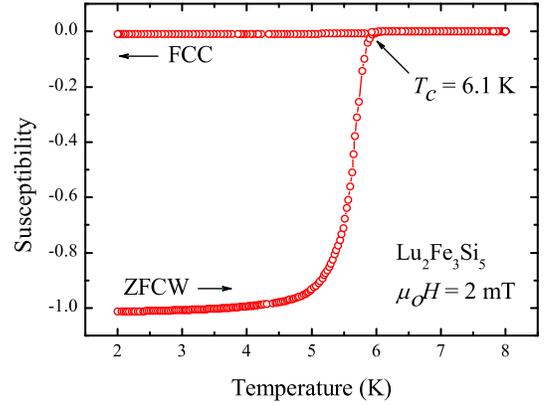} 
\caption{\label{Figure1Biswas} (Color online) The temperature dependence of the dc magnetic susceptibility of Lu$_2$Fe$_3$Si$_5$ measured using both zero-field-cooled warming (ZFCW) and field-cooled cooling (FCC). The diamagnetic susceptibility shows a $T_c$ onset of 6.1~K.}
\end{center}
\end{figure}

A polycrystalline sample of Lu$_2$Fe$_3$Si$_5$ was prepared by melting a stoichiometric mixture of lutetium shot ($99.99\%$), iron granules ($99.999\%$) and silicon pieces ($99.99\%$) in an arc furnace under an argon atmosphere. The as-cast sample was poorly superconducting with a $T_c=4.8$~K and a broad transition. In order to improve these characteristics, it is essential to anneal the as-cast samples at high temperature for a long period of time.~\cite{Tamegai2,Nakajima} The as-cast sample was sealed in a quartz tube under a partial pressure of argon. The sample was then heated at a rate of $200^\circ$C/h to $800^\circ$C, held at this temperature for 48 h, then heated at the same rate  to $1100^\circ$C and held at this temperature for 72 h. The sample was then cooled at $200^\circ$C/h to $800^\circ$C, maintained at this temperature for 72 h, and then finally cooled to room temperature. dc magnetic susceptibility versus temperature ($T$) measurements were performed using a Quantum Design Magnetic Property Measurement System (MPMS) magnetometer. The temperature dependence of the magnetic susceptibility shows that the annealed sample of Lu$_2$Fe$_3$Si$_5$ has a transition temperature, $T_c$ (onset), of 6.1~K [see Fig.~\ref{Figure1Biswas}]. It is the superconducting properties of these annealed samples that are discussed below. Low-temperature specific heat measurements were carried out using a two-tau relaxation method in a Quantum Design Physical Property Measurement System (PPMS) equipped with a $^3$He insert.

The $\mu$SR experiments were performed on the MuSR spectrometer of the ISIS pulsed muon facility. The TF-$\mu$SR experiment was conducted with applied fields between $5$~mT and $60$~mT, which ensured the sample was in the mixed state. The magnetic field was either applied above the superconducting transition and the sample then cooled to base temperature (FC), or the sample was first cooled to base temperature and then the field was applied (ZFC). The MuSR spectrometer comprises 64 detectors. In software, each detector is normalised for the muon decay and reduced to two orthogonal components which are then fitted simultaneously.

The sample was mounted on a silver plate with a circular area of $\sim700$~mm$^2$ and a small amount of diluted GE varnish was added to aid thermal contact. The sample and mount were then inserted into a Oxford Instruments He$^3$ sorbtion cryostat. Any silver exposed to the muon beam gives a non-decaying sine wave.

\begin{figure}[tb]
\begin{center}
\includegraphics[width=0.9\columnwidth]{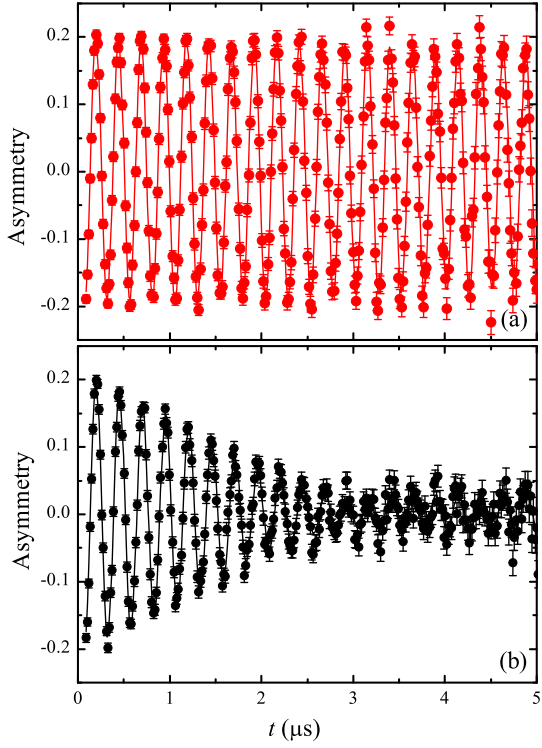}
\caption{\label{Figure2Biswas} (Color online) The transverse-field muon-time spectra (one component) for Lu$_2$Fe$_3$Si$_5$ collected (a) at $T=6.5$~K and (b) at $T=0.3$~K in a magnetic field $\mu_{0}H=30$~mT.}
\end{center}
\end{figure}

\begin{figure}[tb]
\begin{center}
\includegraphics[width=0.9\columnwidth]{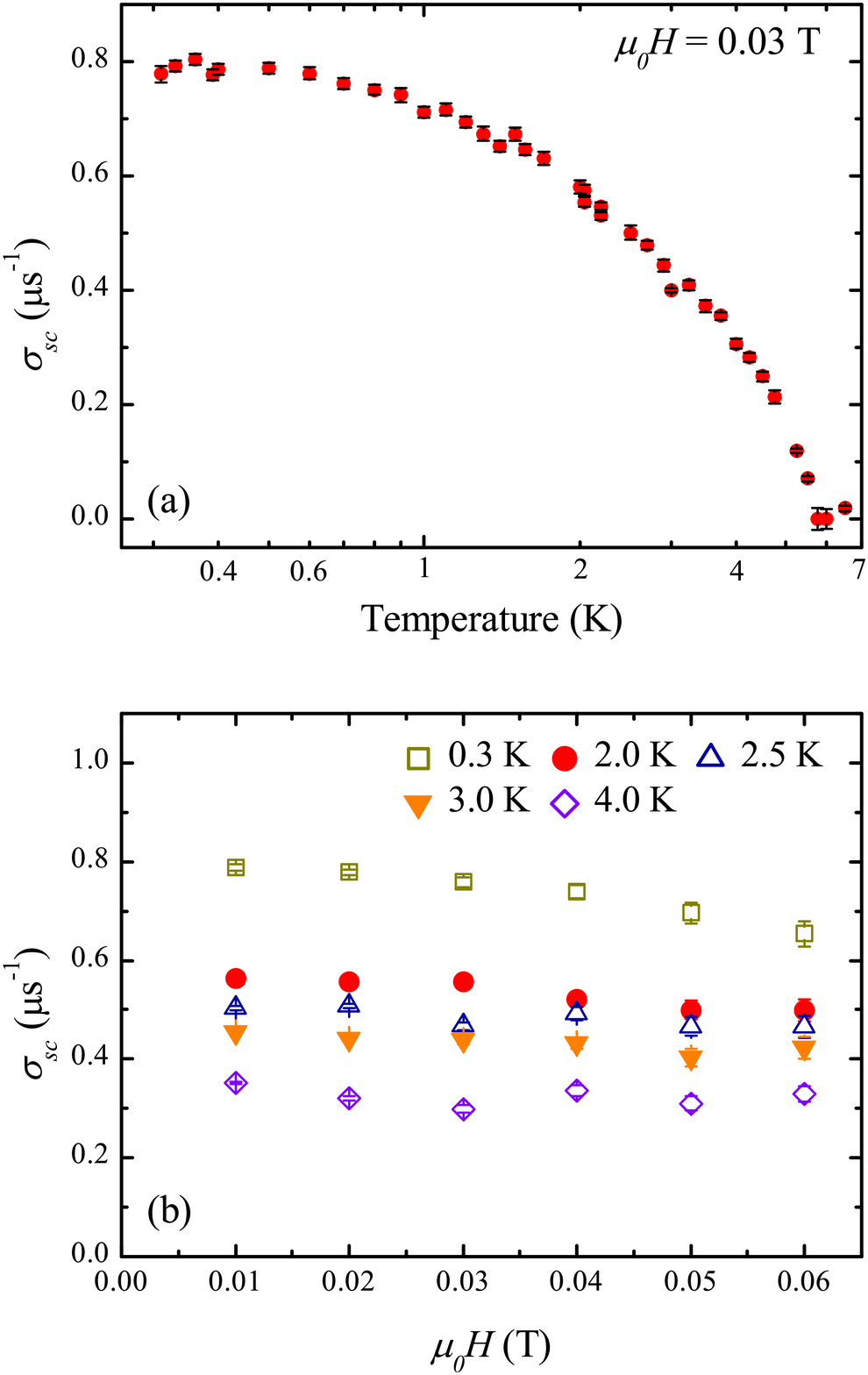}
\caption{\label{Figure3Biswas} (Color online) (a) The temperature dependence (on a log scale) of the superconducting muon spin depolarization rate, $\sigma_{sc}$, collected in an applied magnetic field $\mu_{0}H=30$~mT. (b) Superconducting Gaussian depolarization rate, $\sigma_{sc}$, versus applied magnetic field for Lu$_2$Fe$_3$Si$_5$ collected below $T_c$ at 0.3~K, 2.0~K, 2.5~K, 3.0~K and 4.0~K.}
\end{center}
\end{figure}

TF-$\mu$SR precession signals above and below $T_c=6.1$~K are shown in Figure~\ref{Figure2Biswas}. Above the superconducting transition i.e. in the normal state, the signal decays very slowly, but the decay is relatively fast in the superconducting state due to the inhomogeneous field distribution from the flux-line lattice. We can model these inhomogeneous field distributions using an oscillatory decaying Gaussian function

\begin{eqnarray}
\label{Depolarization_Fit}
G_X(t)=A_{0}\exp\left(-\Lambda t\right)\exp\left(-\sigma^{2}t^{2}\right/2)\cos\left(\omega_{1} t +\phi\right)  \nonumber \\
+A_{1}\cos\left(\omega_{2} t +\phi\right),
\end{eqnarray}
where $\omega_1$ and $\omega_2$ are the frequencies of the muon precession signal and background signal respectively, $\phi$ is the initial phase offset, and $\sigma$ is the Gaussian muon spin relaxation rate. $\sigma$ can also be defined as $\sigma=\left(\sigma^{2}_{sc} + \sigma^{2}_{nm}\right)^{\frac{1}{2}}$, where $\sigma_{sc}$ is the superconducting contribution to the relaxation rate and $\sigma_{nm}$ is the nuclear magnetic dipolar contribution which is assumed to be constant over the entire temperature range. Fig.~\ref{Figure3Biswas}a shows the temperature dependence of $\sigma_{sc}$ obtained in an applied TF of 0.03 T. Fig.~\ref{Figure3Biswas}b presents the magnetic field dependence of $\sigma_{sc}$ collected at different temperatures below the superconducting transition. A deviation in the field dependence of $\sigma_{sc}$  is observed at $40$~mT in $0.3$~K data. A small deviation of $\sigma_{sc}$ is also present at the same field in $2$~K data, whereas it is constant above $2$~K.

The temperature dependence of the London magnetic penetration depth, $\lambda\left(T\right)$ is coupled with the superconducting Gaussian muon-spin depolarization rate, $\sigma_{sc}\left(T\right)$ by the equation

\begin{equation}
\frac{2\sigma^2_{sc}\left(T\right)}{\gamma^{2}_{\mu}}=0.00371\frac{\Phi^{2}_{0}}{\lambda^4\left(T\right)},
\end{equation}
where $\gamma_{\mu}/2\pi=135.5$~MHz/T is the muon gyromagnetic ratio and $\Phi_{0}= 2.068 \times10^{-15}$~Wb is the magnetic flux quantum.~\cite{Sonier, Brandt} $\lambda\left(T\right)$ can be calculated within the local London approximation~\cite{Tinkham,Prozorov} by the following expression

\begin{equation}
\frac{\lambda^{-2}\left(T, \Delta_{0,i}\right)}{\lambda^{-2}\left(0, \Delta_{0,i}\right)}=1+\frac{1}{\pi}\int^{2\pi}_{0}\int^{\infty}_{\Delta_{\left(T,\varphi\right)}}\left(\frac{\partial f}{\partial E}\right)\frac{ EdE d\varphi}{\sqrt{E^2-\Delta_i\left(T,\varphi\right)^2}},
\end{equation}
where $f=\left[1+\exp\left(E/k_BT\right)\right]^{-1}$ is the Fermi function, $\varphi$ is the angle along the Fermi surface, and $\Delta_i\left(T,\varphi\right)=\Delta_{0, i}\delta\left(T/T_c\right)g\left(\varphi\right)$. The temperature dependence of the gap is approximated by the expression $\delta\left(T/T_c\right)=\tanh\left\{1.82\left[1.018\left(T_c/T-1\right)\right]^{0.51}\right\}$ while $g\left(\varphi\right)$ describes the angular dependence of the gap and is replaced by 1 for both an $s$-wave and an  $s+s$-wave gap, and $\left|\cos\left(2\varphi\right)\right|$ for a $d$-wave gap.~\cite{Fang,Errors1}

The temperature dependence of the penetration depth can then be fitted using either a single gap or a two-gap model which are structured on the basis of $\alpha$-model~\cite{Carrington, Padamsee}

\begin{equation}
\label{two_gap}
\frac{\lambda^{-2}\left(T\right)}{\lambda^{-2}\left(0\right)}=\omega_1\frac{\lambda^{-2}\left(T, \Delta_{0,1}\right)}{\lambda^{-2}\left(0,\Delta_{0,1}\right)}+\omega_2\frac{\lambda^{-2}\left(T, \Delta_{0,2}\right)}{\lambda^{-2}\left(0,\Delta_{0,2}\right)},
\end{equation}
where $\lambda^{-2}\left(0\right)$ is the penetration depth at zero-temperature, $\Delta_{0,i}$ is the value of the $i$-th ($i=1$ or 2) superconducting gap at $T=0$~K and $\omega_i$ is a weighting factor with $\omega_1+\omega_2=1$.

\begin{figure}[tb]
\begin{center}
\includegraphics[width=0.9\columnwidth]{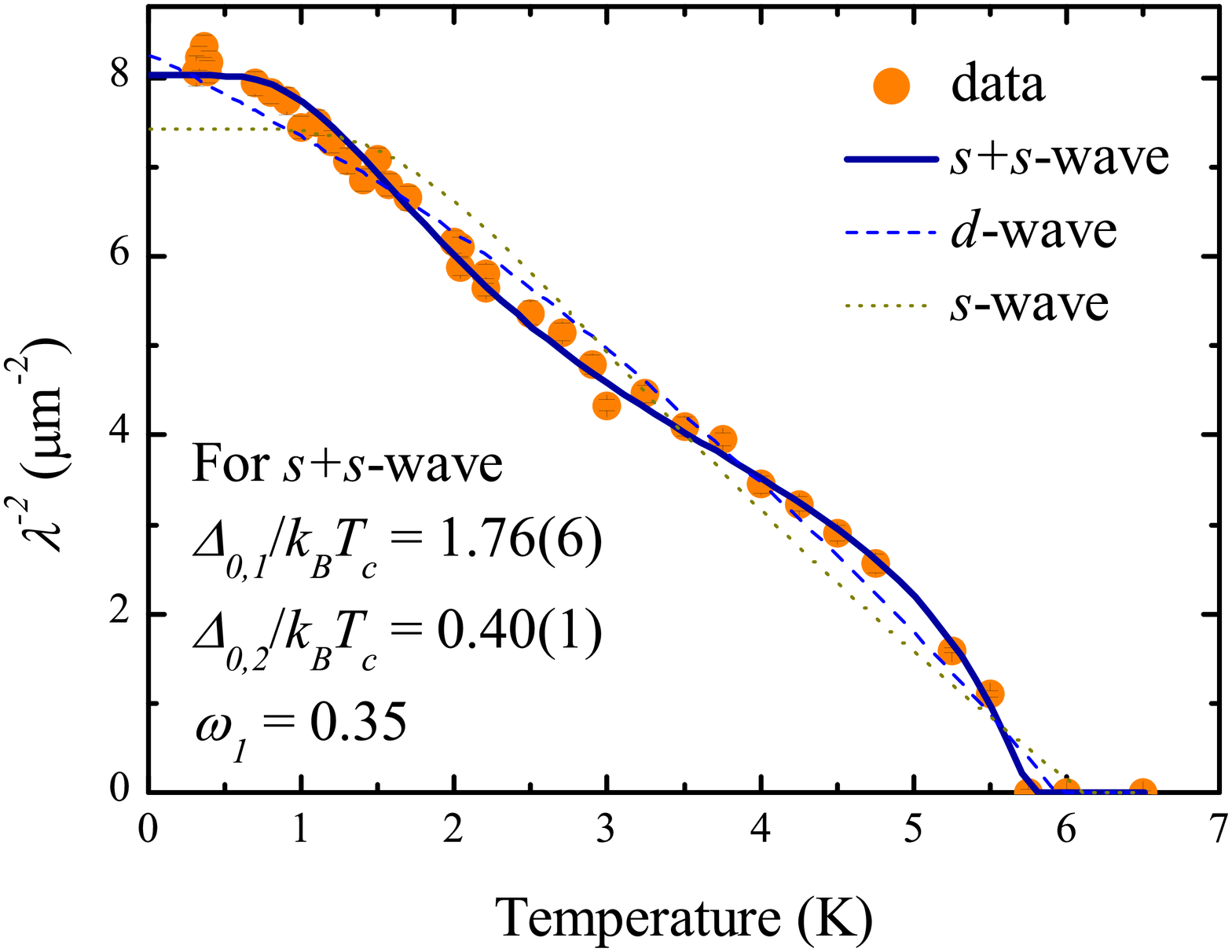}
\caption{\label{Figure4Biswas} (Color online) The temperature dependence of the London penetration depth as a function of temperature for Lu$_2$Fe$_3$Si$_5$. The solid line is a two-gap  $s+s$-wave fit to the data while the dashed and dotted lines represent the $d$-wave and $s$-wave fits respectively.}
\end{center}
\end{figure}

Fits to the data using the three different models are shown in Fig.~\ref{Figure4Biswas}. The fits appear to rule out the $s$-wave and $d$-wave models as possible descriptions for Lu$_2$Fe$_3$Si$_5$ as the $\chi^2$ values for these models are 33.92 and 15.91 respectively. The two-gap  $s+s$-wave model gives a good fit to the data with a $\chi^2$ of 1.94. The two-gap  $s+s$-wave model gives $\Delta_{0,1}/k_{B}T_{c}=1.76(6)$ and $\Delta_{0,2}/{k_{B}T_{c}}=0.40(1)$ with $\omega_1=0.35(1)$. The ratio of larger to the smaller gap, $\frac{\Delta_{0,1}}{\Delta_{0,2}}\approx4.40$, which is consistent with the value 5 obtained by low-temperature specific heat measurement~\cite{Nakajima}, and 3.44 obtained by penetration depth measurement using the tunnel-diode resonator technique~\cite{Gordon} on a single crystal of Lu$_2$Fe$_3$Si$_5$. The magnetic penetration depth at $T=0$~K is found to be $\lambda\left(0\right)=353(1)$~nm. The in-plane penetration depth is 200 nm, obtained by tunnel-diode resonator technique.~\cite{Gordon}

\begin{figure}[tb]
\begin{center}
\includegraphics[width=0.9\columnwidth]{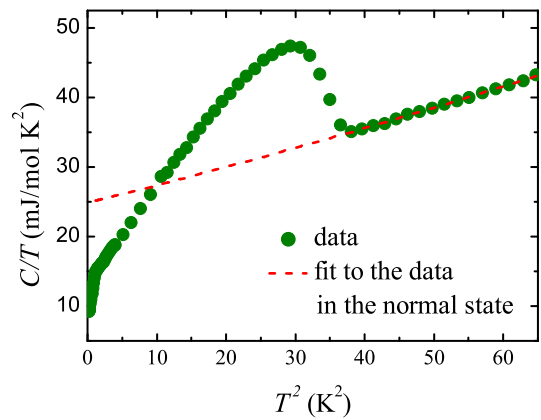}
\caption{\label{Figure5Biswas} (Color online) The specific heat  divided by  temperature ($C/T$) as a function of $T^2$ for Lu$_2$Fe$_3$Si$_5$. The dashed line shows the fit to the data in the normal state.}
\end{center}
\end{figure}

We have also performed low-temperature specific heat measurements on Lu$_2$Fe$_3$Si$_5$ which support the assertion that Lu$_2$Fe$_3$Si$_5$ is a two-gap superconductor. Fig.~\ref{Figure5Biswas} shows the specific heat divided by temperature ($C/T$) as a function of $T^2$ for the same polycrystalline sample of Lu$_2$Fe$_3$Si$_5$ used for the $\mu$SR study. A pronounced jump in the specific heat is observed at 6.1~K which indicates that the sample exhibits bulk superconductivity. The normal state heat capacity has been fitted  up to 12~K by $C=\gamma{T}+\beta{T}^{3}+\alpha{T}^{5}$, where $\gamma{T}$ is the electronic contribution and $\beta{T}^{3}+\alpha{T}^{5}$ represents the lattice contribution to the specific heat. We obtained fitted parameters $\gamma=24.9$~mJ/mol K$^{2}$, $\beta=0.247$~mJ/mol K$^{4}$ and $\alpha=5.38\times{10}^{-4}$~mJ/mol K$^{6}$ which are consistent with the reported values for both polycrystalline~\cite{Vining,Stewart,Tamegai} and single crystal samples.~\cite{Nakajima} We observed a sizeable residual specific heat coefficient, $\gamma_{\circ}=7.21$ mJ/mol K$^2$, at $T=0$~K. Interestingly, a finite residual specific heat coefficient has also been observed in a polycrystalline sample of the same system~\cite{Tamegai} whereas it is absent in data for a single crystal.~\cite{Nakajima}. A similar effect has also been reported in Ba$_{0.6}$K$_{0.4}$Fe$_{2}$As$_{2}$ ($\gamma_{\circ}=7.7$~mJ/mol K$^2$), Ba(Fe$_{1-x}$Co$_x$)$_2$As$_2$ ($\gamma_{\circ}=3.0$ mJ/mol K$^2$) and Ba(Fe$_{0.92}$Co$_{0.08}$)$_2$As$_2$ ($\gamma_{\circ}=3.7$ mJ/mol K$^2$).~\cite{Mu,Mu1,Gofryk} Possible explanations for this residual specific heat coefficient involve pair breaking effects of an unconventional superconductor,~\cite{Kubert} spin glass behavior, or crystallographic defects.~\cite{Gofryk} Given the metallurgy of our polycrystalline sample and the dramatic effects that annealing has on the electronic properties, we suggest that crystallographic defects are the most likely cause of the residual specific heat coefficient in heat capacity data.

\begin{figure}[tb]
\begin{center}
\includegraphics[width=0.9\columnwidth]{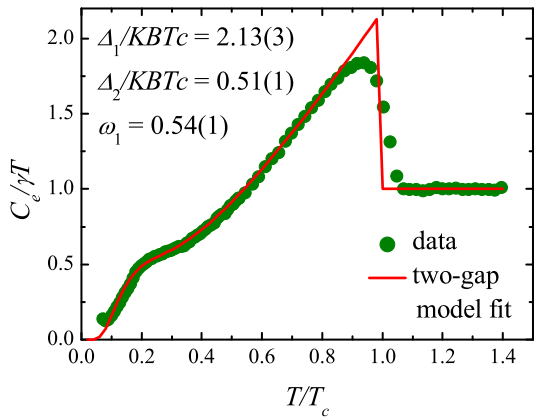}
\caption{\label{Figure6Biswas} (Color online) The temperature dependence of the normalized electronic specific heat as a function of $T/T_c$ for Lu$_2$Fe$_3$Si$_5$. The solid line is a two-gap fit to the data.}
\end{center}
\end{figure}

We have calculated the normalized electronic specific heat, $C_e/\gamma{T}$, by subtracting the lattice contribition from the total heat capacity and then renormalizing by ($\gamma-\gamma_{\circ}$). For more details, see Refs.~\onlinecite{Gofryk} and \onlinecite{Hardy}. Fig.~\ref{Figure6Biswas} shows the temperature dependence of the normalized electronic specific heat, $C_e/\gamma{T}$, for Lu$_2$Fe$_3$Si$_5$ as a function of $T/T_c$. We find two clear anomalies in the data which show that Lu$_2$Fe$_3$Si$_5$ has two energy-gaps. A large jump appears at $T_c$ and a smaller one at $T_c/5$. The value of $C_e/\gamma{T}$ at $T_c$ is found to be 1.13(1) meV, which is much smaller than the BCS value of 1.43 meV but consistent with the value of 1.05 meV measured on a single crystal~\cite{Nakajima} and also agrees well with the reported values for polycrystalline samples.~\cite{Vining,Stewart,Tamegai} To perform a two-gap fit to the $C_e/\gamma{T}$ data in the superconducting state, we use the BCS expressions for the normalized entropy, $S$, and the specific heat

\begin{equation}
\label{two_gap_hc1}
\frac{S}{\gamma_{n}{T_c}}=-\frac{6}{\pi^2}\frac{\Delta_0}{k_{B}T_{c}}\int^{\infty}_{0}[f\ln{f}+(1-f)\ln(1-f)]dy,
\end{equation}

\begin{equation}
\label{two_gap_hc2}
\frac{C}{\gamma_{n}T_{c}}=t\frac{d(S/\gamma_{n}T_{c})}{dt},
\end{equation}
where $t=T/T_c$, $E=[\epsilon^2+\Delta^2(t)]$, and $y=\epsilon/\Delta$. The temperature dependence of the energy gap varies as $\Delta(t)=\Delta_0\delta(t)$, where $\delta(t)$ is the normalized BCS gap.~\cite{Muhlschlegel} The solid line in Fig.~\ref{Figure6Biswas} is a two-gap fit to the data. We obtain two distinct superconducting gaps, $\Delta_1/k_{B}T_{c}=2.13(3)$ and $\Delta_2/k_{B}T_{c}=0.51(1)$. The weighting factor, $\omega_1=0.54$, which is slightly larger than the value obtained from fits to the $\mu$SR data. The good agreement between the experimental heat capacity data and the two-gap model argues in favor of the presence of two distinct superconducting gaps in Lu$_2$Fe$_3$Si$_5$. The ratio of the larger to the smaller gap ($\frac{\Delta_1}{\Delta_2}$) is $4.18$, which is close to the 4.40 obtained from $\mu$SR measurements on the same sample and is also consistent with the published data on Lu$_2$Fe$_3$Si$_5$.~\cite{Vining,Stewart,Tamegai,Nakajima} 

In summary, we have performed a $\mu$SR study on a polycrystalline sample of Lu$_2$Fe$_3$Si$_5$. The temperature dependence of the magnetic penetration depth data was fitted with three different models. A two-gap  $s+s$-wave model provides the best fit to the data. Low-temperature specific heat measurements on the same sample also confirm the presence of two distinct superconducting gaps. The specific heat results can also be reproduced by a two-gap model and support the $\mu$SR results. These results are consistent with other reported data for this system.~\cite{Vining,Stewart,Tamegai,Nakajima,Gordon}

\begin{acknowledgments}
PKB would like to thank the Midlands Physics Alliance Graduate School (MPAGS) for sponsorship. The Quantum Design MPMS magnetometer used in this research was obtained through the Science City Advanced Materials project: Creating and Characterising Next Generation Advanced Materials project, with support from Advantage West Midlands (AWM) and part funded by the European Regional Development Fund (ERDF).     
\end{acknowledgments}

\bibliography{Biswas}

\begin{thebibliography}{25}
\bibitem[{Err()}]{Errors1}
\bibinfo{note}{The normalized $\chi^2$ values, resulting from our least squares
  fits to the temperature dependence of $\lambda^{-2}$ using different models
  for the gap, are used as criteria to determine which model best describes the
  data.}
\end{thebibliography}

\end{document}